\documentclass[10pt,twocolumn,aps,floatfix,citeautoscript,superscriptaddress,prl]{revtex4-2}
\usepackage[english]{babel}

\usepackage{graphicx}
\usepackage[]{subcaption}
\usepackage{natbib}
\usepackage{physics}
\usepackage{xcolor}
\usepackage{tikz}
\usetikzlibrary{arrows.meta}

\usepackage{amsmath, amssymb}
\usepackage{multirow}
\usepackage{xcolor}
\usepackage{pgfplots} 
\usepackage{ulem}
\usepackage{listings}
\usepackage{amsfonts}
\usepackage{comment}
\usepackage[colorlinks=true, allcolors=blue]{hyperref}
\usepackage{ragged2e}
\usepackage{titlesec}

\pgfplotsset{compat=1.18}

\definecolor{codegreen}{rgb}{0,0.6,0}
\definecolor{codegray}{rgb}{0.5,0.5,0.5}
\definecolor{codepurple}{rgb}{0.58,0,0.82}
\definecolor{backcolour}{rgb}{0.95,0.95,0.92}

\newif\ifcomments
\commentsfalse  

\tikzset{dot/.style={draw, thin, circle, fill, outer sep=0pt, inner sep=0pt, minimum width=1mm}}

\titlespacing{\paragraph}{1em}{0em}{0.5em}
\let\originalparagraph\paragraph
\renewcommand{\paragraph}[2][---]{\originalparagraph{#2#1}}

\graphicspath{{./}{figures/}}

\begin{document}

\title{Breakdown of Multipole Expansion in Emergent Electromagnetism}

\author{Tom Ben-Ami}
\email{tom.ben.ami@uni-a.de}
\affiliation{Max-Planck-Institut f\"{u}r Physik komplexer Systeme, N\"{o}thnitzer Stra\ss e 38, Dresden 01187, Germany}
\affiliation{Theoretical Physics III, Center for Electronic Correlations and Magnetism, Institute of Physics, University of Augsburg, D-86135 Augsburg, Germany}

\author{Markus Heyl}
\affiliation{Theoretical Physics III, Center for Electronic Correlations and Magnetism, Institute of Physics, University of Augsburg, D-86135 Augsburg, Germany}
\affiliation{Centre for Advanced Analytics and Predictive Sciences (CAAPS), University of Augsburg, Universitätsstr. 12a, 86159 Augsburg, Germany}

\author{Roderich Moessner}
\affiliation{Max-Planck-Institut f\"{u}r Physik komplexer Systeme, N\"{o}thnitzer Stra\ss e 38, Dresden 01187, Germany}

\begin{abstract}
We find a striking departure from simple electrostatics in a family of two-dimensional lattice models whose effective description takes the form of an emergent electromagnetism: a high-order multipolar charge configuration can induce an electric field characteristic of a \textit{lower-}order multipole. Studying the archetypical hardcore dimer and U(1) spin-1/2 link models, we trace this to a breakdown of  the superposition principle: for dense charge configurations, there can be a nonlinear suppression, or even enhancement, of the far field compared to the linear expectation. Our work establishes that microscopic lattice-scale constraints in models for emergent electromagnetism can change the far-field behaviour to differ from often utilized long-wavelength descriptions in standard effective continuum models. 
\end{abstract}

\date{\today}

\maketitle

\paragraph{Introduction}
In many lattice models in condensed matter physics, effective low-energy descriptions are restricted to manifolds with local constraints. In two dimensions, $U(1)$ constraints often admit a coarse-grained description in terms of a height representation and an emergent gauge theory akin to electromagnetism~\cite{moessner2008quantum,henley2010coulomb}. 
Examples include the $U(1)$ quantum link model~\cite{chandrasekharan97}, hardcore dimer~\cite{rokhsar1988superconductivity, zeng1997zero, AletMisguich2005}, Ising models~\cite{moessner2001ising}, three-colour model on the honeycomb lattice~\cite{huse1992classical}, and more~\cite{kondev1996operator}. 

From the perspective of statistical and soft-matter physics, height representations are particularly relevant for describing surfaces and interfaces. They arise in solid-on-solid models, which describe surface properties~\cite{lapujoulade1994roughening}, and restricted solid-on-solid models~\cite{van1977exactly, andrews1984eight, kim1989growth}. 
Related height descriptions are also central to the statistical mechanics of wetting phenomena~\cite{de1985wetting, privman1988finite}. 
These systems share a common long-wavelength description in terms of a height representation.

More recently, advances in optical lattice experiments~\cite{mil2020scalable, halimeh2025cold}, trapped ion systems~\cite{martinez2016real, banuls2020simulating}, and Rydberg atomic arrays~\cite{surace2020lattice, semeghini2021probing, zeng2025quantum} amongst others have paved the way for the experimental investigation of such $U(1)$ lattice gauge theories, making them a subject of renewed intense research interest~\cite{chakraborty2022disorder, biswas2022scars, karch2026dynamical}.

At long wavelengths, these theories are expected on general grounds to admit an emergent continuum description. For these $U(1)$ models, this takes the familiar structure of electrostatics. One feature of simple electrostatic descriptions is the \textit{multipole expansion}, which states that the potential generated by a local charge distribution takes the form of an angular-harmonic series.  In 2D, this expansion takes the form~\cite{joslin1983multipole}:
\begin{equation} \label{eq:multipole}
    \phi(r, \theta) = a_0 \log(\frac{1}{r}) + \sum_{n=1} \frac{a_n \cos(n\theta) + b_n \sin(n\theta)}{r^n}\, ,
\end{equation}
where the coefficients $a_n, b_n$ encode the multipole moments of the charge configuration. 

\begin{figure*}[!hbt] 
\centering
\includegraphics[width=1.0\linewidth]{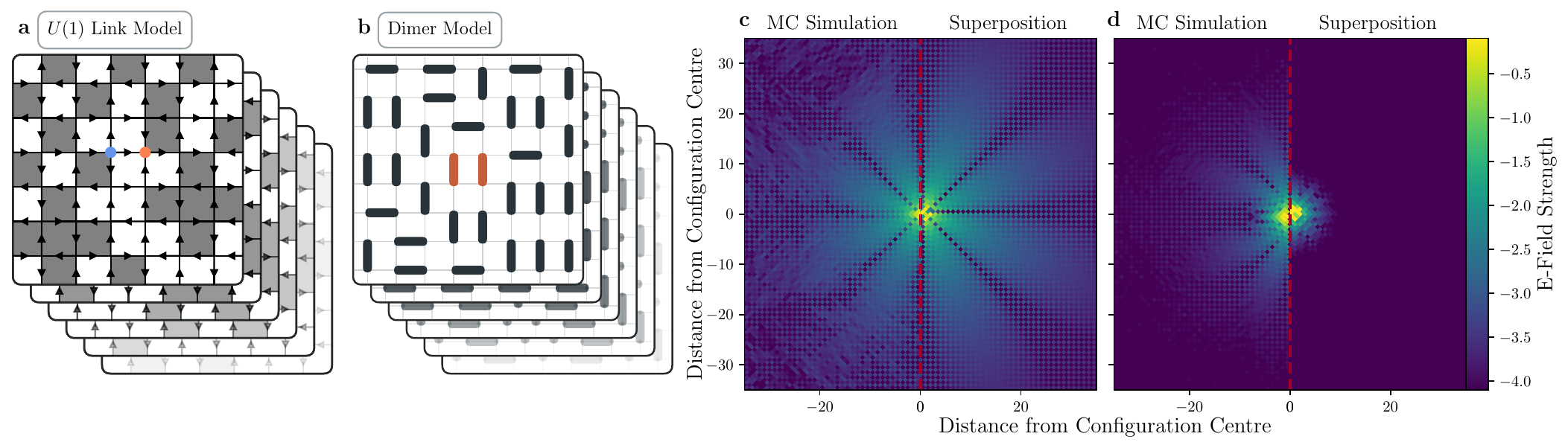}\:
{\phantomsubcaption\label{sfig:linkmod}}
{\phantomsubcaption\label{sfig:dimer}}
{\phantomsubcaption\label{sfig:dipole2d}} 
{\phantomsubcaption\label{sfig:octup2d}}
\caption{\textbf{Multipole Expansion in emergent gauge theories.}
\textbf{a} Schematic of the infinite-temperature ensemble of the $U(1)$ spin-1/2 link model, including two static charges. $\uparrow$ and $\rightarrow$ represent spin $\ket{\uparrow}$ whereas $\downarrow$ and $\leftarrow$ represent spin $\ket{\downarrow}$. Flippable plaquettes are shaded grey, while vertices with net incoming(outgoing) flux are coloured blue(red). 
\textbf{b} Schematic of the infinite-temperature ensemble of the hardcore dimer model on the square lattice; the highlighted dimers represent frozen dimers.
\textbf{c} Monte Carlo simulation (left) and linear-response expectation (right) for the link model for a dipolar configuration. 
\textbf{d} Monte Carlo simulation (left) and linear-response expectation (right) for the link model for an octupolar configuration. 
}
\label{fig:decays_dipole}
\end{figure*}

Here, we show that in constrained lattice models this correspondence to the continuum descriptions fails. Specifically, we find that the multipole expansion breaks down in two archetypical models of emergent gauge theories, namely the $U(1)$ spin-$1/2$ link model (Fig.~\ref{sfig:linkmod}) and the hardcore dimer model (HDM, Fig.~\ref{sfig:dimer}).

\begin{figure*}[!htb] 
\centering
\includegraphics[width=0.95\linewidth]{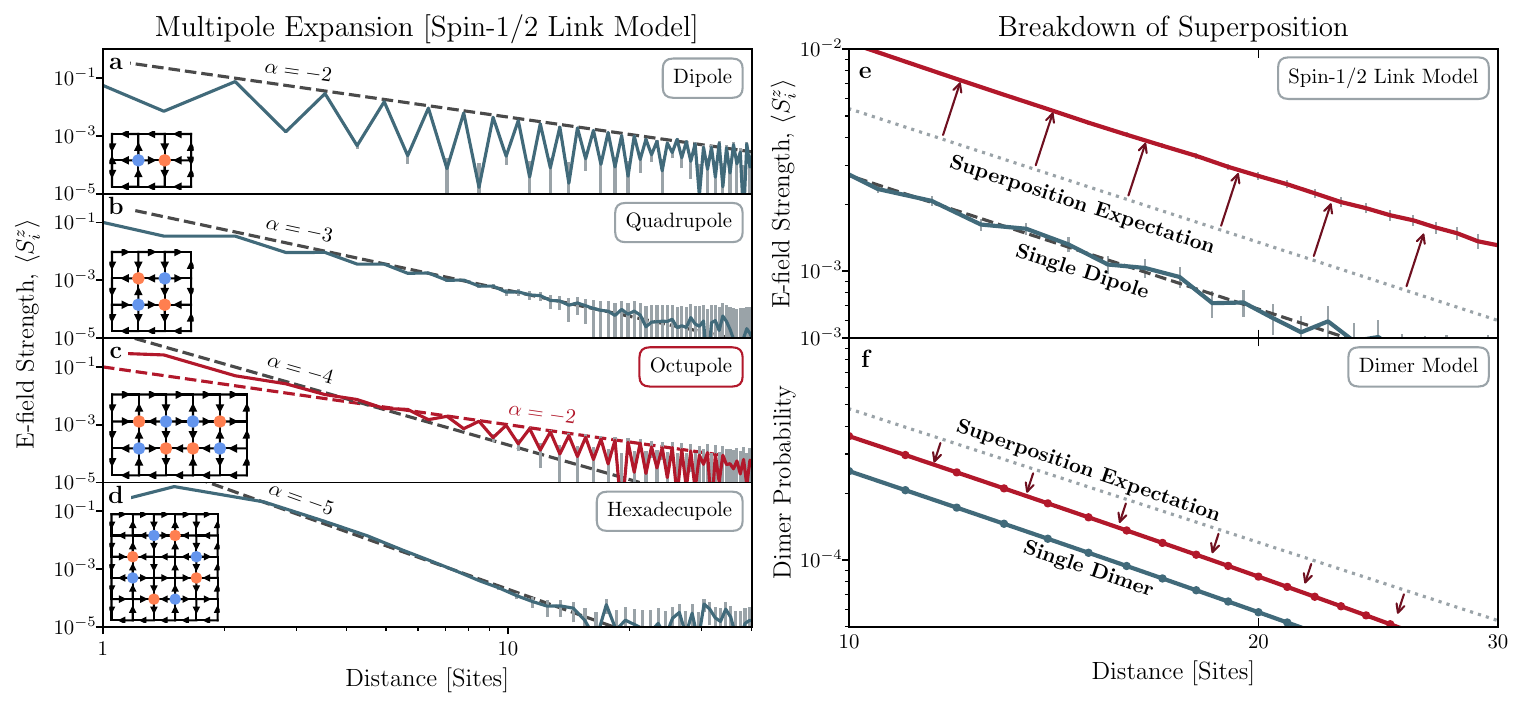}\:
{\phantomsubcaption\label{sfig:dipole}}
{\phantomsubcaption\label{sfig:quadrupole}}
{\phantomsubcaption\label{sfig:octupole}}
{\phantomsubcaption\label{sfig:hexadecupole}}
{\phantomsubcaption\label{sfig:dbldipole}}
{\phantomsubcaption\label{sfig:dimerdipole}}
\caption{\textbf{Breakdown of multipole expansion and superposition principle.}
\textbf{a-d} Multipole expansion in the $U(1)$ spin-$1/2$ link model. Monte Carlo simulations are performed on $120\times120$ systems, with the multipole configurations placed at the centre. The exact charge configuration is shown in each respective inset. The multipole expansion breaks down at octupolar order, as shown in \textbf{c}, where the asymptotic response becomes dipolar.
\textbf{e} Breakdown of superposition for two dipoles in the link model. 
\textbf{f} Breakdown of superposition for two dipoles in the HDM. Dimer occupation probabilities are obtained using the Kasteleyn matrix, from a $200\times 180$ system.}
\label{fig:multipoles}
\end{figure*}

Both display clear and simple departures from linear electrostatics and the continuum description.
We find that the far-field response of a pure multipole charge configuration can acquire {\it lower}-order multipole components, even when these are absent in the bare charge distribution (Fig.~\ref{sfig:octup2d}).
Just as strikingly, in these systems the superposition principle itself breaks down --- combining two dipolar excitations does not yield the linear sum expected from electrostatics.
In the link model, this produces a nonlinear {\it enhancement} of the far-field response (Fig.~\ref{sfig:dbldipole}) that preserves its dipolar character, whereas the HDM exhibits a suppression relative to the linear expectation (Fig.~\ref{sfig:dimerdipole}).

We trace this non-linear behaviour to the constraints the gauge degree of freedom are subject to at the lattice scale.
These strongly distort the near-field flux structure and consequently the long-range response of the systems. 
These effects represent a form of UV-IR mixing, as microscopic short-wavelength constraints directly modify the emergent long-wavelength response.
In such models, microscopic lattice scale constraints thus generate a non-linear long-wavelength response absent from the naive formulation of the coarse-grained descriptions. 

\paragraph{Models} \label{sec:models}
We consider two constrained lattice models that realise emergent gauge theory descriptions with strongly truncated local degrees of freedom for the gauge fields. 
Despite their distinct microscopic origins, both admit coarse-grained descriptions resembling electrostatics~\cite{nienhuis1984critical}.
This provides a natural setting to investigate departures from linear coarse-grained descriptions in such emergent gauge theories of strongly constrained models.
We consider the uniform classical statistical ensemble of allowed configurations.

The 2D spin-$1/2$ $U(1)$ link model (Fig.~\ref{sfig:linkmod}) is an Abelian lattice gauge theory frequently used as an effective model for lattice electrodynamics~\cite{chandrasekharan97,halimeh2025cold}, in which the gauge fields reside on the links of the lattice. Unlike its high-energy counterparts, the local degree of freedom on each link is finite, a spin-$1/2$, reflecting the available degrees of freedom in microscopic condensed matter models and quantum simulators.
Throughout our work we consider the classical statistical mechanics of the pure gauge sector without dynamical matter.

The allowed configurations of such gauge fields are constrained by Gauss’ law, which enforces local flux conservation at every vertex through the gauge operator
\begin{equation} \label{eq:gauge}
    G_{\vb{r}} = \sum_\mu \qty(S^z_{\vb{r},\mu} - S^z_{\vb{r}-\mu, \mu}).
\end{equation}
This measures the net flux in/out of the vertex at position $\vb{r}$. For the square lattice, the eigenvalues of $G_{\vb{r}}$ for spin-$1/2$ links are then $q_{\vb{r}} \in \qty{-2, -1, 0, 1, 2}$.
The gauge operator labels each unique configuration of $\qty{q_{\vb{r}}}$ as a distinct superselection sector. These sectors can be interpreted as a static charge background, embedded in an otherwise dynamic gauge field (see Fig.~\ref{sfig:linkmod} for a sector with a dipole-like charge pair at its centre).
As a result, the link model provides a natural minimal setting for investigating the phenomenology of constrained emergent gauge theories.

As a microscopically distinct constrained system with a related coarse-grained description, we additionally consider the hardcore dimer model on the square lattice.
A configuration consists of each site covered by exactly one dimer, so that allowed configurations, $C_d$, are close‑packed `perfect coverings' of the lattice (Fig.~\ref{sfig:dimer}).

The classical system is exactly solvable via the Kasteleyn matrix, an oriented antisymmetric adjacency matrix whose Pfaffian gives the partition function~\cite{kasteleyn1961}. On the square lattice, correlation functions of dimer-dimer occupations reduce to determinants of the inverse Kasteleyn matrix. In the thermodynamic limit these decay algebraically as $r^{-2}$~\cite{fisher1961statistical, fisher1963statistical}.

Similarly to the link model, the HDM admits a height representation~\cite{nienhuis1984critical, henley1997relaxation}, providing a coarse-grained field description that connects local constraints to an emergent gauge theory. Although the microscopic details differ substantially between the two models, both generate similar effective descriptions, allowing us to probe different behaviours within this class of constrained models.

\paragraph{Breakdown of the multipole expansion} \label{sec:breakdown} 
We first investigate whether the spin-$1/2$ link model admits a conventional multipole description. To do so, we construct background charge configurations whose lowest nonvanishing moments are the dipole $(n^2)$, quadrupole ($n^4$), octupole ($n^6$, sometimes termed sextupoles), and hexadecapole ($n^8$) moments, respectively. We then compare the long-distance gauge field response to the n{a}\"{i}ve expectations of Eq.~\ref{eq:multipole}. 
We sample the link model using cluster Monte Carlo simulations, as detailed in the End Matter.

We begin with the dipole, the leading contribution for a charge-neutral configuration.
We find that the leading far-field behaviour reproduces the conventional multipolar prediction, and the electric field $\abs{\vb{E}(r)}$ scales with the real-space distance to the dipole, $r$, as $\abs{\vb{E}(r)} \propto r^{-2}$. It exhibits an additional subleading contribution decaying as $\abs{\vb{E}(r)}\propto r^{-3}$ (Fig.~\ref{sfig:dipole}). 
This is captured naturally within a linear-response treatment based on the height representation, detailed in the End Matter.
The analytic linear-response solution clearly reproduces the Monte Carlo simulations, including the subleading $r^{-3}$ correction, as demonstrated in Figure~\ref{sfig:dipole2d} where the two are compared. 

We furthermore find that for both the quadrupole and hexadecupole moments, the resulting fields exhibit the expected $r^{-3}$ and $r^{-5}$ far-field decays respectively, as predicted by Eq.~\ref{eq:multipole} (Fig.~\ref{sfig:quadrupole}, \ref{sfig:hexadecupole}). In these cases, the emergent gauge-field response remains well described by the coarse-grained description and linear response. 

In stark contrast, the octupolar configuration violates the multipole picture, as observed in Figure~\ref{sfig:octup2d}. 
Rather than exhibiting the expected $r^{-4}$ far-field behaviour, the response is dominated by an induced \textit{dipolar} component, despite the absence of a dipole moment in the bare charge configuration (Fig.~\ref{sfig:octupole}): a lower-order response is generated even though its corresponding moment is absent in the bare charge configuration.
Such leading-order contribution cannot be explained solely through the effect of lattice anisotropy on the Laplacian, as the anisotropic correction to the Laplacian only introduce subleading terms (see SM for further details on the effect of lattice anisotropy).

The breakdown of the multipole expansion, through the emergence of a lower-order moment, demonstrates that such strongly constrained emergent gauge fields can diverge from linear electrostatics.
Notably, while at long distances lattice effects are conventionally expected to become inconsequential~\cite{michta2021asymptotic}, we find that in the link model the long-distance response is directly tied to the short-distance lattice effects. 

\paragraph{Breakdown of superposition} \label{sec:ddip}
We trace the breakdown of the multipole expansion to a more general failure of the {superposition principle} in emergent gauge systems with strong constraints.
To demonstrate this, we investigate the gauge-field response generated by a pair of neighbouring dipoles. 

In the spin-$1/2$ link model, the far-field response of the gauge field generated by two adjacent dipoles deviates strongly from the prediction of linear superposition (Fig.~\ref{sfig:dbldipole}).
The far-field response retains its dipolar structure, both in terms of the decay exponent and the angular dependence, but has an amplitude markedly {\it larger} than that obtained by linearly superposing the fields of the two isolated dipoles. Adding  opposing dipoles to generate an octupole then does not cancel this enhancement, instead yielding the apparent net dipole moment visible in the far field.

\begin{figure}[!hbt] 
\centering
\includegraphics[width=1.0\linewidth]{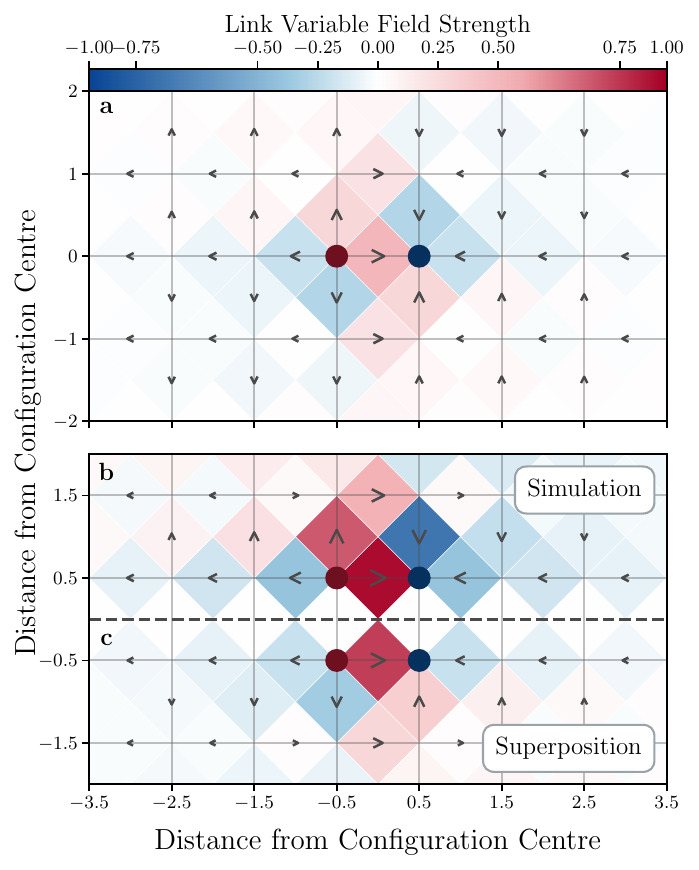}\:
{\phantomsubcaption\label{sfig:NF1}}
{\phantomsubcaption\label{sfig:NF2}}
{\phantomsubcaption\label{sfig:NF3}}
\caption{\textbf{Dipole near-field.} 
Close-up of the link model gauge field around: \textbf{a} -- a single dipole and \textbf{b} -- two adjacent dipoles. The link variables are represented by the diamond-shaped tiles.
\textbf{c} shows the field expected from superposition of the configuration in \textbf{a}.
The sign indicates the direction of the link spin, also represented by the arrow direction, while the colour encodes its magnitude. Charge locations are shown as red and blue circles. 
}
\label{fig:nearfield}
\end{figure}

To gain an intuition for the mechanism underpinning this nonlinearity, consider how the limited degree of freedom -- just the two spin states -- on each link restricts the amount of flux that it can accommodate. 
Let us identify the origin of this effect through examining the gauge near-field response around the charge configuration (Fig.~\ref{fig:nearfield}).
Compared to a single dipole, the spin links surrounding the two dipoles are significantly closer to the limits of their allowed maximal field strength.
In particular, for this charge configuration, the vertical links carry zero flux while the central links connecting each dipole are saturated. The local flux distribution is therefore strongly limited from organising independently around each dipole. The near-field flux distribution is thus strongly distorted from a sum of two dipoles, which in turn results in the enhancement of the far-field response. 

To determine whether this breakdown of superposition is specific to the link model or instead arises generically from strong local constraints, we turn to the HDM on the square lattice, which has the added benefit of being exactly soluble. We examine the dimer-dimer correlations as a function of distance. 
For the square-lattice dimer model, the connected correlation function decays asymptotically as~\cite{fisher1963statistical}:
\begin{equation}
    C_{ij}(r) = \expval{n_i n_{j + \vb{r}}} - \expval{n_i} \expval{n_{j + \vb{r}}} \propto \frac{1}{r^2}.
\end{equation}
Within the height representation, a single frozen dimer corresponds to two opposite charges located at its endpoints.

Using the Kasteleyn matrix formalism, we evaluate the corresponding two- and three-dimer correlation functions exactly~\cite{kasteleyn1961} (See SM for details).
We find that for the two-dipole configuration, the induced long-distance dimer-occupation response still shows a dipolar decay, but is \textit{suppressed} in comparison to the superposition expectation (Fig.~\ref{sfig:dimerdipole}). 
This behaviour contrasts with the link model, where an enhancement was seen. The same underlying phenomenon ---  local constraints modify the emergent gauge-field response to multiple sources --- thus appears generic to this family of models.

We further develop a phenomenological description of this nonlinear response by relating the local plaquette activity observed in the Monte Carlo chains of the link model to an effective permittivity, as detailed in the Supplementary. 

\paragraph{Discussion \& Outlook}
\label{sec:disc}

Emergent gauge theories are often interpreted through the lens of continuum electrostatics, where microscopic details such as lattice anisotropy are expected to affect only subleading corrections at sufficiently large scales~\cite{michta2021asymptotic}. 
Our work shows that constrained emergent gauge theories can instead retain a leading-order dependence on their microscopic constraints.
Even within classical statistical mechanics, fundamental features of continuum electrostatics, such as linear superposition and the conventional multipole expansion, need not hold in such emergent theories. This is important as such models are prevalent in condensed matter physics, from frustrated magnets~\cite{henley2010coulomb, balents2010spin} to ultracold atomic setups~\cite{wiese2013ultracold}. 

Importantly, the breakdown of the coarse-grained linear response description is not restricted to a single microscopic realization. The link model and HDM exhibit qualitatively different nonlinear responses, demonstrating that the collective response of constrained emergent gauge systems depends on the microscopic structure of the constraint. While the link model exhibits an enhancement of the far field and the HDM displays suppression, both reveal the same underlying principle: local constraints alter the effective response of the emergent gauge field.

The same physics can also be viewed through the height representation, making it also particularly relevant to models in soft matter.
In particular, the corresponding height representation is closely related to restricted solid-on-solid models.
Our results therefore suggest that nonlinear response in restricted solid-on-solid models need not originate from spatial disorder or heterogeneity, but can instead emerge from the effects of local configurational constraints.

Beyond equilibrium and electrostatics, these observations also raise questions about the dynamics of constrained gauge systems. Looking at plaquette activity, constraint-induced cooperativity is visible in the Monte Carlo sampling dynamics, where regions of enhanced or suppressed collective plaquette activity develop around static charges (see SM). These correlations suggest that the nonlinear static response is accompanied by cooperative dynamics, providing a possible route toward understanding the dynamical consequences of strong local constraints.

\begin{acknowledgements}

\textbf{Acknowledgements} --- 
This work was supported in part by the Deutsche Forschungsgemeinschaft via the cluster of excellence ctd.qmat (EXC 2147, Project-ID 390858490) and  FOR 5522 (Project-ID No. 499180199).
The authors gratefully acknowledge the resources on the LiCCA HPC cluster of the University of Augsburg, co-funded by the Deutsche Forschungsgemeinschaft (DFG, German Research Foundation) – Project-ID 499211671.
\end{acknowledgements}

\textit{Data availability —} The data to generate all figures in this letter is available in Zenodo \citep{data_repo}.

\appendix

\section{End Matter}

\begin{figure}[!hbt]
\centering
\includegraphics[width=0.95\linewidth]{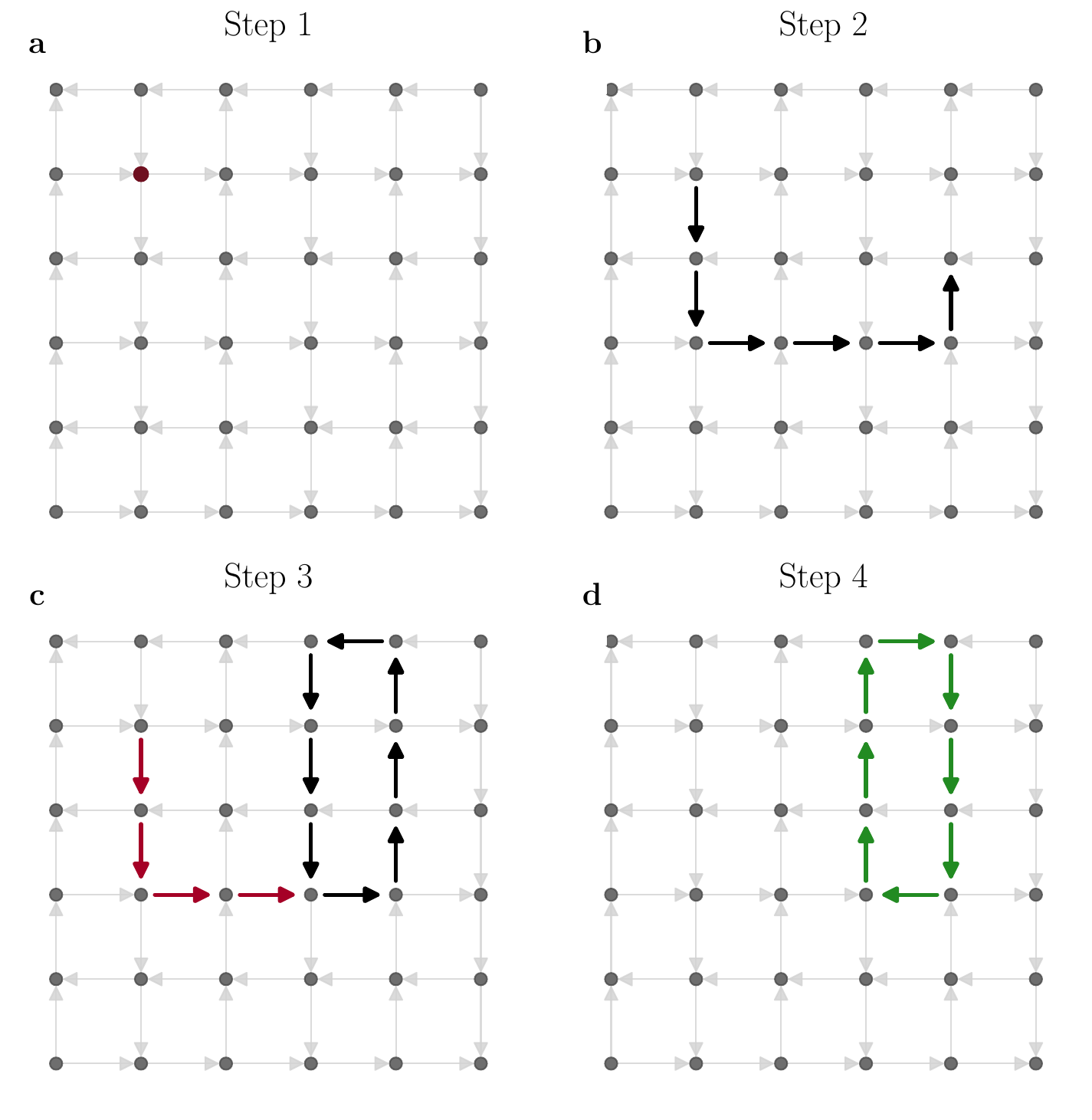}\:
{\phantomsubcaption\label{sfig:Cstp1}}
{\phantomsubcaption\label{sfig:Cstp2}}
{\phantomsubcaption\label{sfig:Cstp3}}
{\phantomsubcaption\label{sfig:Cstp4}}
\caption{\textbf{Short-loop Monte Carlo algorithm.} \textbf{a} A random vertex is chosen in the system. \textbf{b} A directed random walk through the system is performed. \textbf{c} Once the walker intersects itself, the `tail' is discarded, leaving a closed flippable loop. \textbf{d} The closed loop is flipped, completing the update.}
\label{fig:CMC}
\end{figure}

\paragraph{Methods} \label{app:method} 
We compute the background-charge-induced gauge field response of the link model using classical Monte Carlo simulations. 
We employ two update schemes: local plaquette flips and global short-loop updates (Fig.~\ref{fig:CMC}).
The local update consists of flipping a plaquette subject to the gauge constraint:
\begin{equation}
\makebox[\columnwidth][c]{%
  \begin{tikzpicture}[baseline=-0.5ex]
    \node (p) at (0,0) {\tikz[
  baseline=-0.5ex,
  scale=0.45,
  >={Stealth[length=5pt,width=5pt]}
]{
  \fill (0,0) circle (2.0pt);
  \fill (2,0) circle (2.0pt);
  \fill (2,2) circle (2.0pt);
  \fill (0,2) circle (2.0pt);

  \draw
    (0,0) -- (2,0) -- (2,2) -- (0,2) -- cycle;

  \draw[->] (0.35,0) -- (1.15,0);
  \draw[->] (2,0.35) -- (2,1.15);
  \draw[->] (1.65,2) -- (0.85,2);
  \draw[->] (0,1.65) -- (0,0.85);
}};
    \node (c) at (2.0,0) {\tikz[
  baseline=-0.5ex,
  scale=0.45,
  >={Stealth[length=5pt,width=5pt]}
]{
  \fill (0,0) circle (2.0pt);
  \fill (2,0) circle (2.0pt);
  \fill (2,2) circle (2.0pt);
  \fill (0,2) circle (2.0pt);

  \draw
    (0,0) -- (2,0) -- (2,2) -- (0,2) -- cycle;

  \draw[->] (1.65,0) -- (0.82,0);
  \draw[->] (2,1.65) -- (2,0.82);
  \draw[->] (0.35,2) -- (1.18,2);
  \draw[->] (0,0.35) -- (0,1.18);
}};
    \draw[<->, >=Stealth] (p.east) -- (c.west);
  \end{tikzpicture}%
}
\label{eq:H_LM}
\end{equation}
generated by the plaquette flip operator $U_{\square} = S^+_{\vb{r}, i} S^+_{\vb{r}+i, j} S^-_{\vb{r}+j, i} S^-_{\vb{r}, j}$, where $S^{+/-}$ are spin raising/lowering operators. 
The operator acts only on flippable plaquettes, for which the links form a closed loop, such that the flip preserves the gauge constraint (see Fig.~\ref{sfig:linkmod}).

The short-loop update flips closed loops of varying sizes in the system while preserving the gauge constraint.
Loops are constructed through a random walk on the directed graph, which terminates once the walker reaches a previously visited vertex (see Fig.~\ref{fig:CMC}). The remaining tail is then discarded from the update, leaving a closed, flippable loop.
All simulations are conducted at infinite temperature. 
Every simulation consists of $1000$ independent chains, each of length $10^{10}$ steps.
A burn-in period of $t_{burn}\approx 2000\; \tau_{\text{int}}$ is applied for each chain prior to taking measurements.
Measurements are then taken at intervals chosen to produce a set of statistically uncorrelated snapshots of the ensemble.

The gauge field response is then obtained from the ensemble average, computed as an equal-weight average over all measurements from all independent chains,
\begin{equation}
    \expval{S_{x,y}^m} = \frac{1}{N_\text{meas}} \sum_{n=1}^{N_\text{meas}} (S_{x,y}^m)_n.
\end{equation}
Here, $m=h/v$ is the horizontal/vertical orientation of the link, while $x, y$ specify the position of the vertex to the left and bottom of the link. 
Both update schemes conserve the background-charge configuration. We therefore initialise all chains within the same superselection sector, with a background-charge configuration chosen to realise the desired multipole moments of Eq.~\ref{eq:multipole} at the centre of the system.

\paragraph{Linear Response}
The additional $r^{-3}$ correction observed in the link model dipole configuration arises naturally within a linear-response treatment based on the height representation. 
The creation of a pair of particle-antiparticle charges in the link model is equivalent to flipping spins along a path connecting the two charges, while preserving the gauge constraint at the vertices along the interior of the path. To linear order, we approximate this as a weak bias applied along the path within the charge-free superselection sector of the link model, 
\begin{equation}
    H = H_{LM} - \mu\!\sum_{(i,j; n)\in \text{path}} h_{i,j}^n(S^z)^n_{i,j},
\end{equation}
where $\mu$ is small and $h_{i,j}^n=\pm 1$ is a uniform-magnitude bias whose sign on the link $(i, j; n)$ is chosen to favour the required spin flip for creating the particle-antiparticle pair. Here $i,j$ label the lattice site, while $n\in \{1,2\}$ denotes the link orientation (horizontal/vertical).

Assuming a charge-free background such that $\expval{S^z_{i,j}} = 0$, linear response yields,
\begin{equation} \label{eq:spin_lr}
    \expval{(S^{z})^{n}_{i,j}} = \beta\mu \sum_{(l, k; m) \in \text{path}} \expval{(S^z)_{i,j}^n(S^z)_{l,k}^m}. 
\end{equation}
The spin-spin correlation functions are then found using the height representation~\citep{moessner04planar}:
\begin{equation} \label{eq:LR_cart}
    \begin{aligned}
        \expval{S^1_{0,0}S^1_{x, y}} & = \Xi \frac{y^2 - x^2}{\qty(x^2 + y^2)^2} + \frac{\Upsilon}{\qty(x^2 + y^2)^{3/2}},\\
        \expval{S^2_{0,0}S^2_{x, y}} & = \Xi \frac{ x^2 - y^2}{\qty(x^2 + y^2)^2} + \frac{\Upsilon}{\qty(x^2 + y^2)^{3/2}},\\
        \expval{S^1_{0,0}S^2_{x, y}} & = \Xi \frac{ 2xy}{\qty(x^2 + y^2)^2} - \frac{\Upsilon}{\qty(x^2 + y^2)^{3/2}}.\\
    \end{aligned}
\end{equation}
The correlators consist of two distinct contributions: an $r^{-2}$ term with $\cos(2\theta)$ or $\sin(2\theta)$ angular dependence, and an angle-independent $r^{-3}$ contribution whose amplitude $\Upsilon$ depends on the charge configuration.  

For a dipole constructed from nearest-neighbour vertices, the path consists of a single link in Eq.~\ref{eq:spin_lr}. 
In this case, the induced long-distance field is directly proportional to the corresponding spin-spin correlator.
This is in agreement with the Monte Carlo simulations (see Fig.~\ref{fig:decays_dipole}).
Notably, this $r^{-3}$ scaling arises from correlations within the gauge field and is independent of the underlying multipole configuration.

\bibliography{references}

\onecolumngrid

\pagebreak

\appendix


\onecolumngrid

\begin{center}
{\large\bfseries Supplemental Material: \\ Breakdown of Multipole Expansion in Emergent Electromagnetism\par}
\vspace{3mm}



\end{center}


\section{Effect of lattice anisotropy} 
Lattice anisotropy arises from the mismatch between the continuous symmetries of the continuum theory and the discrete symmetries of the underlying lattice, which introduces corrections to the continuum Green's function that reflect the underlying lattice symmetry.
We examine whether these corrections for the square lattice can generate a leading-order contribution. Consider the Green's function
\begin{equation}
    G(\vb{r}) = \int_{-\pi}^{\pi} \int_{-\pi}^{\pi} \frac{dk_x dk_y}{4\pi^2} \frac{e^{i\vb{k} \vdot\vb{r}}}{\mathcal{L}(\vb{k})}  
\end{equation}
where $\mathcal{L}(\vb{k}) = 4 - 2\cos k_x - 2\cos k_y$. 
For small wavevectors, we expand with respect to $k$ as $\cos k_i \approx 1 - k_i^2/ 2 + k^4_i/24 - \mathcal{O}(k^6_i)$. Thus,
\begin{equation}
\begin{split}
    \mathcal{L}(\vb{k}) &= 4 - 2( 1 - k_x^2/ 2 + k^4_x/24) - 2( 1 - k_y^2/ 2 + k^4_y/24) \\
    &= k_x^2 + k_y^2 - \frac{k_x^4 + k_y^4}{12} + \mathcal{O}(k^6) \\
    &= \vb{k}^2 - \frac{k_x^4 + k_y^4}{12} + \mathcal{O}(k^6).
\end{split}
\end{equation}
Therefore, the leading anisotropic contribution appears at fourth order in the wavevector. Introducing polar coordinates, this becomes:
\begin{equation}
\begin{split}
    \mathcal{L}(\vb{k}) &= k^2 - \frac{k^4}{12} \qty( \cos^4\theta + \sin^4\theta) \\
    &= k^2 - \frac{k^4}{12} ( \frac{\cos(4\theta)}{4} + \frac{3}{4}) \\
    &= k^2 \qty( 1 - \frac{k^2}{16} - \frac{k^2\cos(4\theta)}{48}).
\end{split}
\end{equation}
Taking the inverse as 
\begin{equation}
\frac{1}{\mathcal{L}(\vb{k})} = \frac{1}{k^2} \frac{1}{1 - \frac{k^2}{16} - \frac{k^2\cos(4\theta)}{48}},
\end{equation}
and expanding again for small $k$ gives
\begin{equation}
\frac{1}{\mathcal{L}(\vb{k})} \approx \frac{1}{k^2} (1 + \frac{k^2}{16} + \frac{k^2 \cos(4\theta)}{48} + \mathcal{O}(k^4)).
\end{equation}
Fourier transforming the individual terms gives
\begin{equation}
\begin{split}
    \int \frac{d^2k}{4\pi^2} \frac{e^{i\vb{k}\vdot\vb{r}}}{k^2} &= -\frac{1}{2\pi} \ln r, \\
    \int \frac{d^2k}{4\pi^2} e^{i\vb{k}\vdot\vb{r}} &= \delta(r), \\
    \int \frac{d^2k}{4\pi^2} e^{i\vb{k}\vdot\vb{r}} \cos(4\theta) &= \frac{2}{\pi}\frac{\cos(4\phi)}{r^2}.
\end{split}
\end{equation}
The first term gives the familiar logarithmic continuum Green's function, the second gives a contact term, and the third provides the leading square-lattice anisotropic correction.
\begin{equation}
    G(\vb{r}) = -\frac{1}{2\pi} \ln r + \frac{1}{16}\delta(r) + \frac{1}{24\pi r^2}\cos(4\phi) + ...
\end{equation}
The anisotropic correction decays as $r^{-2}$ and carries a fourfold angular dependence imposed by the square lattice. Applying the octupolar operator $O_{ijk}\partial_i\partial_j\partial_k$ to the isotropic logarithmic contribution produces a term scaling as $r^{-3}$, whereas acting on the anisotropic correction produces a sub-leading contribution scaling as $r^{-5}$. Thus, lattice anisotropy cannot generate a leading-order dipolar contribution. Lattice effects alone are therefore insufficient to account for the observed leading-order dipole term.

\section{A Phenomenological Picture}
\subsection{Activity}

The emergence of a dipolar response from the octupolar configuration suggests that the conventional multipole description fails to describe the effects of static charges on the gauge field under a homogeneous background. To understand the origin of this induced dipolar response, we introduce a phenomenological description based on the notion of \textit{activity}.

We define the activity $A_{ij}$ of a plaquette at position $(i, j)$ as the number of times that a plaquette is flipped during a single Metropolis Monte Carlo trajectory. Therefore, the activity measures how strongly different regions in the constrained gauge field participate in the dynamics. 
To obtain a relative measure of activity across the system, we normalise by the mean activity:
\begin{equation}
    \bar{A}_{ij} = \frac{A_{ij}}{\frac{1}{L^2}\sum_{i, j}A_{ij}}.
\end{equation}
We then use $\bar{A}_{ij}$ to define a phenomenological, spatially dependent, effective permittivity by averaging over the four plaquettes adjacent to each vertex,
\begin{equation}
\tilde{\varepsilon}_{i,j} = 1/4  \!\!\!\!\!\!\!\!\!\!\!\! \sum_{\square\in\text{neighbours}\ (i, j)} \!\!\!\!\!\!\!\!\!\bar{A}_{\square}.
\end{equation}
This effective permittivity provides a phenomenological representation of the spatially varying dynamical response induced by the charge configuration and its effect on the equilibrium response field.
Spatial variations in the effective permittivity lead to a change in the effective environment experienced by the charges, thereby requiring corrections to the conventional multipole expansion. 

\begin{figure}[!htb] 
\centering
\includegraphics[width=0.5\linewidth]{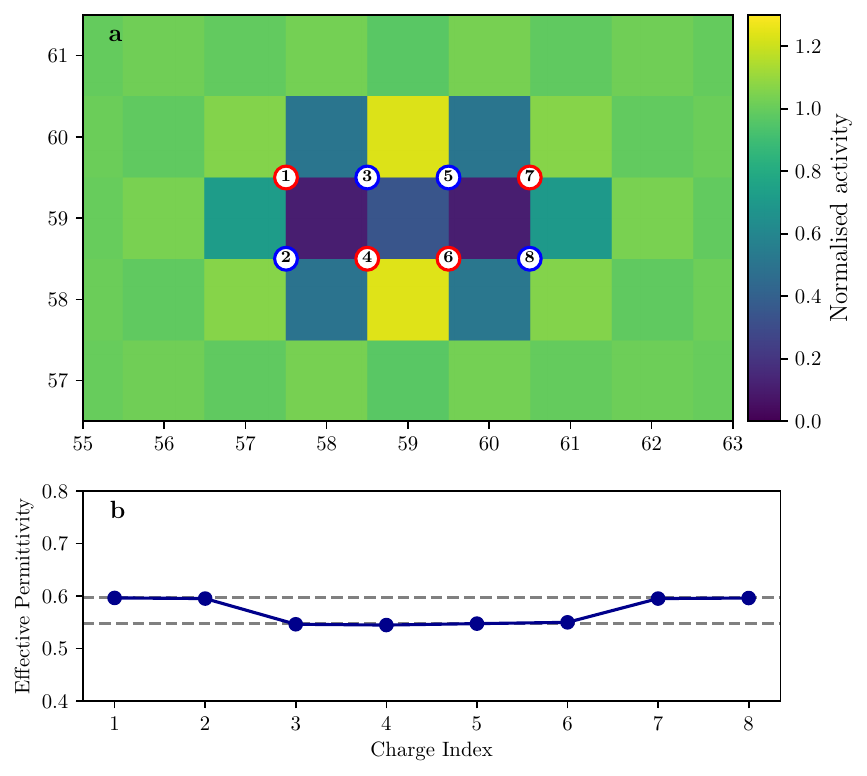}\:
{\phantomsubcaption\label{sfig:activity}}
{\phantomsubcaption\label{sfig:permitivity}}
\caption{\textbf{Activity measure of the octupole configuration.} \textbf{a} Normalised activity around the octupole charges in the link model. \textbf{b} Effective permittivity around each charge. The indices correspond to the indices seen in \textbf{a}. }
\label{fig:activity}
\end{figure}

For the octupolar configuration, we find that the activity, and therefore the effective permittivity, develop pronounced variation around the charge configuration (see Fig.~\ref{fig:activity}). 
Specifically, the spatial variation of the average activity defines two regions of distinct effective permittivity, separated by effective `emergent' interfaces. Consequently, individual charges experience different local environments.
Within this phenomenological picture, these emergent interfaces generate corrections to the conventional multipole structure and give rise to a dipolar contribution to the far field despite the absence of a microscopic dipole moment.
The resulting correction terms can be derived analytically within this framework, as presented below.

\subsection{Octupole with an Interface} \label{app:interface}
The analysis of the octupole charge configuration shows that the two outer dipoles, indexed by $\qty{(1,5), (4,8)}$ in figure~\ref{fig:activity}, exhibit higher activity than the two inner dipoles, indexed by $\qty{(2, 6), (3, 7)}$. 

\begin{figure}[!ht]
\centering
\includegraphics[width=1.0\linewidth]{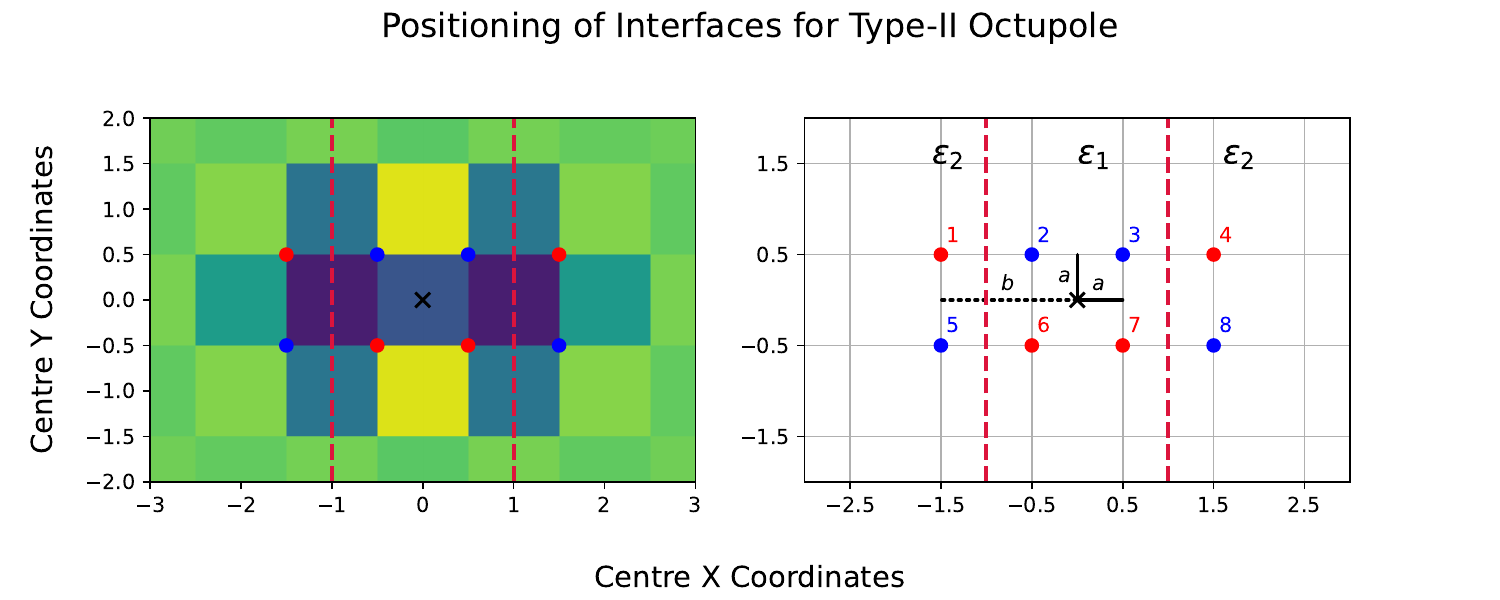}
\caption{Placement of interfaces between regions of different permittivity in the octupole. Right - the length scales $a,b$ used for calculating the multipole terms are shown. Left - Positioning of interfaces on the activity map of the system.}
\label{fig:act_int}
\end{figure}

Motivated by the difference in activity discussed above, we introduce two interfaces separating regions with distinct relative permittivities, $\varepsilon_1, \varepsilon_2$. We position the interfaces and divide the octupole as shown in figure \ref{fig:act_int}.
We then recompute the potential at a point \((x,y)\) far from the charge configuration, accounting for the spatially varying permittivity. The resulting potential is:
\begin{align}
\Phi(x, y)
  = \frac{q}{2\pi \varepsilon_0}
     &\bigg[
        \frac{1}{\varepsilon_1}
        \big(
            \log|\vb{r}_6| - \log|\vb{r}_2|
            + \log|\vb{r}_7| - \log|\vb{r}_3|
        \big) \nonumber\\[6pt]
  &+ \frac{1}{\varepsilon_2}
        \big(
            \log|\vb{r}_1| - \log|\vb{r}_5|
            + \log|\vb{r}_4| - \log|\vb{r}_8|
        \big)
     \bigg]
\end{align}
where $\vb{r}_i = (x - x_i, y-y_i)$ with $(x_i, y_i)$ denoting the position of charge $i$ as indexed in figure~\ref{fig:act_int}.
Rewriting using a single relative permittivity $\varepsilon=\frac{\varepsilon_1}{\varepsilon_2}$,
\begin{align}
\Phi(r,\theta)
  = \frac{q}{2\pi \varepsilon_1 \varepsilon_0}
     & \bigg\{
        \big[
            \log|r_6| - \log|r_2|
        \big]
        \nonumber 
    + \big[
            \log|r_7| - \log|r_3|
        \big]
        \nonumber\\[4pt]
  &+ \varepsilon
        \big[
            \log|r_1| - \log|r_5|
            + \log|r_4| - \log|r_8|
        \big]
     \bigg\}
\end{align}

Finally, assuming $r >> a, b$ we find that the correction multipole expansion yields:
\begin{equation}
    \begin{split}
        \phi^{(0)}(r, \theta) & = 0 \\
        \phi^{(1)}(r, \theta) & = \frac{4\sin(\theta)}{r}\qty[1 - \varepsilon] \\
        \phi^{(2)}(r, \theta) & = 0, \\
        \phi^{(3)}(r, \theta) & = \frac{4a\sin(\theta)}{r^3}\qty[ 3\cos^2(\theta)\qty(a^2-b^2) + \qty(a^2\sin^2(\theta) + 3b^2\cos^2(\theta))\qty(1-\varepsilon)].
    \end{split}
\end{equation} 
At large distances, the leading contribution is therefore dipolar. The octupolar term $\phi^{(3)}$ includes a $\varepsilon$-independent contribution, attributed to the original octupolar term without interfaces, and an $\varepsilon$-dependent enhancement term arising from the interfaces. 
For \(\varepsilon=1\), the \(r^{-1}\) term vanishes, recovering the original naive octupolar field in the homogeneous case.

\section{Dimer Model Kasteleyn Matrix Correlator}
\label{app:wick}
To analyse correlations in the hardcore dimer model, we use the Kasteleyn matrix, $M$, to evaluate dimer correlation functions.
Using the corresponding Wick expansion, we evaluate the three-dimer correlator and its connected contribution, corresponding to the two-dipole configuration.
We consider correlation functions between one or two dimers located at the centre of the system and an asymptotically distant 'test dimer'. The test dimer allows us to probe the far-field response generated by the central dimer configuration.

\begin{figure}[!hbt] 
\centering
\includegraphics[width=1.0\linewidth]{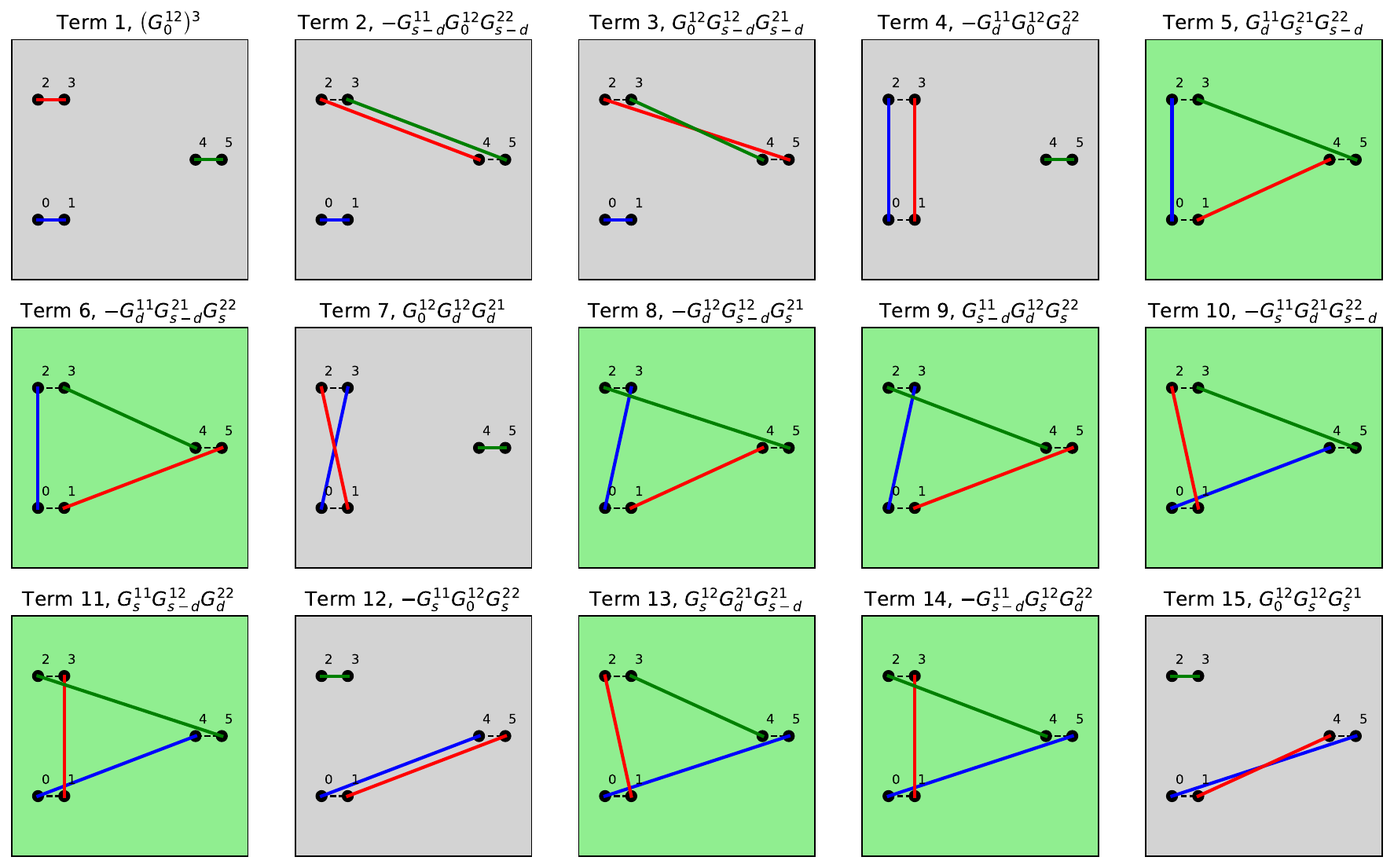}\:
\caption{\textbf{Wick terms for three dimers.} All possible Wick contractions for the three-dimer correlator. The terms highlighted in green correspond to the connected contributions that generate the nonlinear correction to the far-field response.
}
\label{fig:Wick}
\end{figure}

Following the approach described in Refs.~\cite{fisher1963statistical, fendley2002classical}, the dimer correlation functions can be expressed in terms of Grassmann variables. In this representation, the two-point Green's function is given by
\begin{equation}
    \expval{\psi_i \psi_j} = \frac{1}{Z}\int \qty[\mathcal{D}\psi] \psi_i \psi_j \exp(S).
\end{equation}
Using a two-site unit cell, we re-index as $i \rightarrow \ell, \vb{x}$ where $\ell \in \{1, 2\}$ labels the two sublattices of the bipartite lattice, and $\vb{x}$ denotes the position. The action is then 
$S = \frac{1}{2} \sum_{\alpha, \vb{x}} \sum_{\beta, \vb{y}} \psi_{\alpha, \vb{x}} M^{\alpha \beta}_{\vb{x} \vb{y}} \psi_{\beta, \vb{y}} $.
The corresponding Green's functions are then:
\begin{equation}
    \begin{split}
        \expval{\psi_{1, \vb{x}} \psi_{1, \vb{x}+\vb{r}}} &= G^{11}_{r}, \\
        \expval{\psi_{2, \vb{x}} \psi_{2, \vb{x}+\vb{r}}} &= G^{22}_{r}, \\
        \expval{\psi_{1, \vb{x}} \psi_{2, \vb{x}+\vb{r}}} &= G^{12}_{r}, \\
        \expval{\psi_{2, \vb{x}} \psi_{1, \vb{x}+\vb{r}}} &= G^{21}_{r}, \\
    \end{split}
\end{equation}
We consider three dimers at positions:
\begin{equation}
    \begin{split}
        C_a &= (1, \vb{0})-(2, \vb{0}), \\
        C_b &= (1, \vb{d})-(2, \vb{d}), \\
        C_c &= (1, \vb{s})-(2, \vb{s}).
    \end{split}
\end{equation}
Here, $C_a$ and $C_b$ form the central two-dimer configuration, while $C_c$ denotes the asymptotically distant test dimer.
We assume $\abs{\vb{s}} \gg \abs{\vb{d}}$; the corresponding geometry is shown in figure~\ref{fig:Wick}.
The sublattice labels determine the orientation of each dimer: a dimer is oriented from sublattice $1$ to sublattice $2$ or vice versa. Two adjacent dimers with the same orientation have aligned dipoles and retain a net dipole moment, whereas dimers with opposite orientations have oppositely aligned dipoles and form a quadrupole.
The three-dimer correlator is then given by
\begin{equation}
\expval{C_a C_b C_c} = 
\expval{\psi_{1, \vb{0}} \psi_{2, \vb{0}} 
\psi_{1, \vb{d}} \psi_{2, \vb{d}}
\psi_{1, \vb{s}} \psi_{2, \vb{s}}}.
\end{equation}
Applying Wick's theorem, the three-dimer correlator can be expressed as a sum of products of two-point Green's functions. The complete set of Wick contractions is shown in Fig.~\ref{fig:Wick}.

We finally evaluate the full three-dimer correlator numerically using the Kasteleyn matrix as a function of $\vb{s}$. From the resulting Wick contributions, we isolate the terms relevant to the far-field response of the central two-dimer configuration and use their asymptotic behaviour to extract the corresponding dipolar contribution.

\pagebreak

\section{Cooperative Plaquette Dynamics}
\begin{figure*}[!hbt]
\centering
\includegraphics[width=1.0\linewidth]{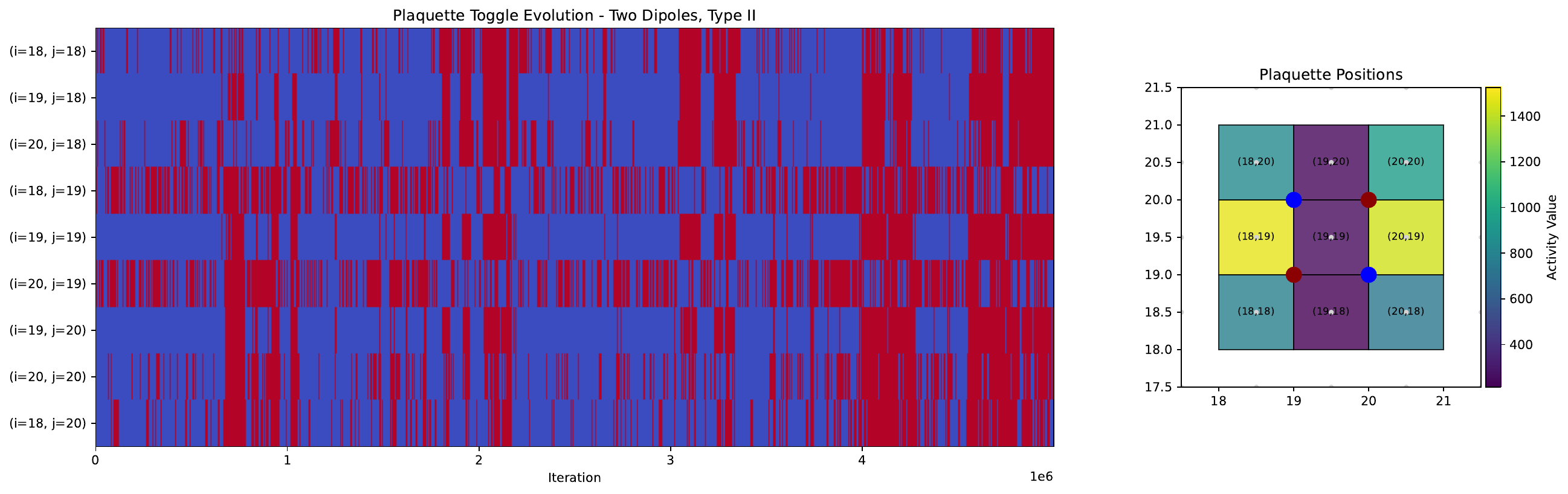}\:
\caption{\textbf{Plaquette Cooperativity around a pair of Dipoles.} The activity of the neighbouring plaquettes is tracked according to their orientation, with clockwise plaquettes shown in blue and counter-clockwise plaquettes shown in red. The time traces illustrate the cooperative behaviour of neighbouring plaquettes, which tend to become active or inactive together during a single Monte Carlo trajectory.}
\label{fig:cooper}
\end{figure*}

We examine the local dynamics around the charges. Specifically, we analyse the time traces of the nine plaquettes neighbouring the double-dipole charge configuration and track their activity during a single Monte Carlo trajectory (Fig.~\ref{fig:cooper}).
 
We find that plaquettes around the charges tend to become active or inactive together, indicating cooperative behaviour. This is a direct result of the system's strong constraints, whereby flipping a plaquette strongly influences the flippability of its neighbours. 

\end{document}